# Photoionization-induced broadband dispersive wave generated in an Ar-filled hollow-core photonic crystal fiber

**Jianhua Fu** [1,2,†], **Yifei Chen** [1,2,†], **Zhiyuan Huang** [1,2,*], **Fei Yu** [3,4], **Dakun Wu** [4], **Jinyu Pan** [1,2], **Cheng Zhang** [4], **Ding Wang** [1], **Meng Pang** [1,4] **and Yuxin Leng** [1,4,*]

[1] State Key Laboratory of High Field Laser Physics and CAS Center for Excellence in Ultra-intense Laser Science, Shanghai Institute of Optics and Fine Mechanics (SIOM), Chinese Academy of Sciences (CAS), Shanghai 201800, China
[2] Center of Materials Science and Optoelectronics Engineering, University of Chinese Academy of Sciences, Beijing 100049, China
[3] R&D Center of High Power Laser Components, Shanghai Institute of Optics and Fine Mechanics, Chinese Academy of Sciences, Shanghai 201800, China
[4] Hangzhou Institute for Advanced Study, Chinese Academy of Sciences, Hangzhou 310024, China
[*] Corresponding author: huangzhiyuan@siom.ac.cn; lengyuxin@mail.siom.ac.cn
[†] These authors contributed equally to this work.

**Abstract:** The resonance band in hollow-core photonic crystal fiber (HC-PCF), while leading to high-loss region in the fiber transmission spectrum, has been successfully used for generating phase-matched dispersive wave (DW). Here, we report that the spectral width of the resonance-induced DW can be largely broadened due to plasma-driven blueshifting soliton. In the experiment, we observed that in a short length of Ar-filled single-ring HC-PCF the soliton self-compression and photoionization effects caused a strong spectral blueshift of the pump pulse, changing the phase-matching condition of the DW emission process. Therefore, broadening of DW spectrum to the longer-wavelength side was obtained with several spectral peaks, which correspond to the generation of DW at different positions along the fiber. In the simulation, we used super-Gauss windows with different central wavelengths to filter out these DW spectral peaks, and studied the time-domain characteristics of these peaks respectively using Fourier transform method. The simulation results verified that these multiple-peaks on the DW spectrum have different delays in the time domain, agreeing well with our theoretical prediction. Remarkably, we found that the whole time-domain DW trace can be compressed to ~29 fs using proper chirp compensation. The experimental and numerical results reported here provide some insight into the resonance-induced DW generation process in gas-filled HC-PCFs, they could also pave the way to ultrafast pulse generation using DW-emission mechanism.

**Keywords:** Hollow-core photonic crystal fiber; soliton; photoionization; dispersive wave

## 1. Introduction

Phase-matched dispersive wave (DW), arising from a nonlinear energy transfer from a self-compressed soliton, has attracted great research interests in the past few decades [1-4]. In particular, the low-loss broadband-guiding gas-filled hollow-core photonic crystal fibers (HC-PCFs), as ideal platforms, are widely used for studying the efficient generation of tunable phase-matched DW [5-7]. The phase-matched DW can not only be generated in a wide wavelength range such as ultraviolet (UV) [8] and mid-infrared (MIR) spectral regions [9], but its wavelength can also be effectively tuned, which makes this kind of light sources have many applications, especially in ultrafast spectroscopy.

In the gas-filled HC-PCFs, the combined effects of self-compressed soliton and high-order dispersion can result in phase-matched DW generated in the UV region [5,7]. As early as ten years ago, Joly et al. demonstrated the efficient emission of DW in the deep-UV region by using an Ar-filled kagomé-style HC-PCF [8]. The generated bright spatially coherent deep-UV laser source is tunable from 200 to 320 nm through varying the pulse energy and gas pressure. Belli et al. used a short length of hydrogen-filled kagomé-style HC-PCF to develop the generation of vacuum-UV (VUV) to near-infrared (NIR) supercontinuum [10]. The experimental result showed that a strong VUV DW emission generated at 182 nm on the trailing edge of the pulse, and it also proved that kagomé-style HC-PCF works well in the VUV spectral region. Simultaneously, Ermolov et al. reported on the generation of a three-octave-wide supercontinuum from VUV to NIR region in a He-filled kagomé-style HC-PCF [11]. The VUV pulses generated through DW emission is tunable from 120 to 200 nm, with efficiencies >1% and VUV pulse energy >50 nJ. It should be pointed out that in the strong-field regime, the plasma caused by gas ionization modifies the fiber dispersion, allowing phase-matched DW generation in the MIR region. Novoa et al. first predicted the generation of MIR DW, and established a new model to explain it well [12]. Köttig et al. first experimentally demonstrated the existence of MIR DW and realized a 4.7-octave-wide supercontinuum from 180 nm to 4.7 μm, with up to 1.7 W of total average power [9].

The phase-matched DW can not only be generated in the UV and MIR spectral regions, but also in the visible and NIR regions. Sollapur et al. experimentally reported on the generation of a three-octave-wide supercontinuum from 200 nm to 1.7 μm at an output energy of ~23 μJ in a Kr-filled HC-PCF [13]. Simulations showed that the spectra generated in the visible and NIR resonance bands are closely related to the emission of the phase-matched DW. Tani et al. further proved that the narrow-band spectral peaks generated in the resonance bands are based on the phase-matched DW emission due to the anti-crossing dispersion [14]. Recently, Meng et al. demonstrated the generation of NIR DW in the resonance bands of an Ar-filled kagomé-style HC-PCF [15]. In our recent experiments [16,17], we reported the high-efficiency emission of phase-matched DW in the visible spectral region using a He-filled single-ring (SR) HC-PCF. As the input pulse energy increases, the central wavelength of the plasma-driven blueshifting soliton (BS) [18,19] is close to the resonance band of the fiber, high-efficiency energy transfer from the pump light to the DW can be triggered.

In this work, we demonstrated in the experiments that the photoionization-induced broadband DW in a 25-cm-long Ar-filled SR HC-PCF. In a certain pulse energy region, we observed that soliton self-compression of the input pulse results in a spectral expansion that overlaps with resonant DW frequencies, leading to a narrow-band DW spectral peak in the first resonance band of the SR HC-PCF. At high pulse energy levels, we observed that the plasma-driven BS can further excite multiple DW peaks, which leads to the broadening of DW spectrum to the longer-wavelength side. In addition, we theoretically investigated the time-domain characteristics of these spectral peaks filtered by the super-Gauss windows using Fourier transform method, and compressed the DW pulses to ~29 fs through suitable dispersion compensation.

## 2. Experimental Set-up

The experimental set-up is shown in Fig. 1. The 800 nm, ~45 fs, 0.3 mJ pulses from a commercial Ti:Sapphire laser system were coupled into a 1-m-long hollow-core fiber (HCF) by using a concave mirror with a focal length of 1 m. The HCF has a core diameter of 250 μm and was placed in a gas cell filled with 212 mbar Ar. The spectrum of the pulses after propagating the Ar-filled HCF was broadened due to self-phase modulation (SPM), and the energy transmission was measured to be ~70%. Several pairs of chirped mirrors (CMs) were not only used to compensate the output pulses from the gas-filled HCF, but also to pre-compensate the dispersion introduced by some optical elements, including half-wave plate (HWP), wire grid polarizer (WGP), and even plano-convex lens (PCL) and the window at the input port of the second gas cell. Then the compressed pulses were launched into a 25-cm-long SR HC-PCF filled with 1.4 bar Ar using a coated PCL with a focal length of 0.1 m. The insert indicates the scanning electron micrograph (SEM) of the SR HC-PCF that has a core diameter of 24 μm and a wall thickness of ~0.26 μm.

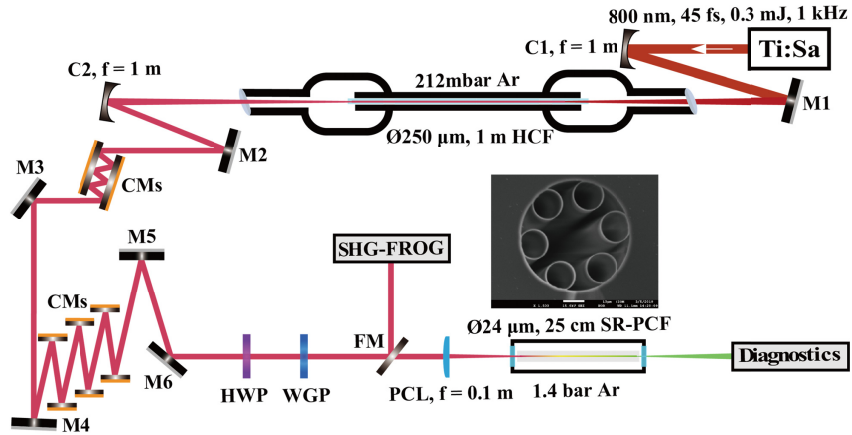

**Figure 1.** Schematic of the experimental set-up. M1-M6, silver mirrors; C1-C2, concave mirrors; CMs, chirped mirrors; HWP, half-wave plate; WGP, wire grid polarizer; FM, flipping mirror; PCL, plano-convex lens. The inset in the set-up represents SEM of the SR-PCF with a core diameter of 24 μm and a wall thickness of ~0.26 μm.

In Fig. 2(a), we plot the simulated (green solid line) and measured (purple solid line) fiber losses of the fundamental optical mode $HE_{11}$ of the SR HC-PCF. The simulated fiber loss was calculated by the bouncing ray (BR) model [20]. Figure 2(b) shows the dispersion curve of Ar-filled SR HC-PCF at gas pressure of 1.4 bar calculated by the fully analytical model created by Zeisberger and Schmidt (ZS) [21]. In both Figs. 2(a) and 2(b), the cyan bars point out the first resonance band of the SR HC-PCF. In the experiments, we used a home-built second-harmonic-generation frequency-resolved optical gating (SHG-FROG) to measure the input pulses in front of the PCL. The retrieved trace with an error of 0.6%, as shown in Fig. 2(d), agrees well with the measured trace in Fig. 2(c). The retrieved temporal and spectral intensities of the input pulses and their phase are plotted in Figs. 2(e) and 2(f). The measured input pulse duration is ~18.7 fs [full width at half-maximum (FWHM)].

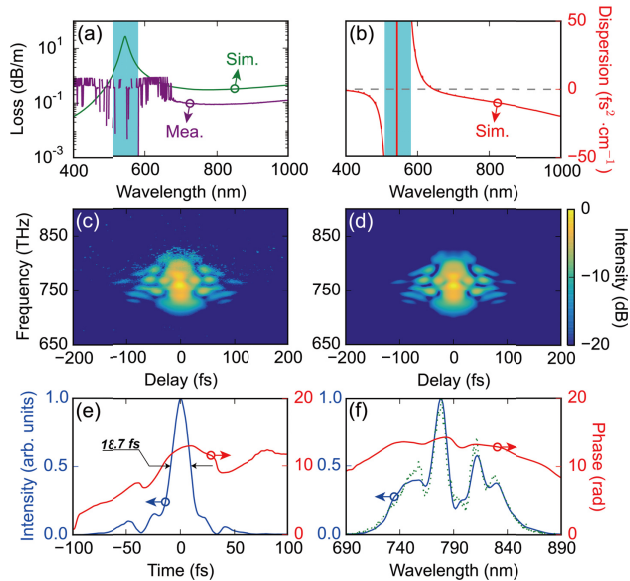

**Figure 2.** (a) Simulated (green solid line) and measured (purple solid line) fiber losses of the fundamental mode $HE_{11}$ of the SR HC-PCF. (b) Simulated dispersion (red solid line) of the SR HC-PCF filled with 1.4-bar Ar gas. The simulated fiber loss and dispersion are calculated through the BR and ZS models, respectively. In both (a) and (b), the cyan bars show the first resonant spectral region of the SR HC-PCF. (c) and (d) Measured and retrieved SHG-FROG traces of the pulses in the front of the focusing lens. (e) and (f) Retrieved temporal and spectral profiles (blue solid lines) and the corresponding phase (red solid lines). The green dotted line in panel (f) represents the measured reference spectrum.

## 3. Experimental Results and Analysis

As show in Fig. 3(a), we measured the spectral evolutions after propagating a 25-cm-long Ar-filled SR HC-PCF (24 μm core diameter and ~0.26 μm wall thickness) at gas pressure of 1.4 bar as a function of input pulse energy from 0.5 μJ to 3.2 μJ. Moreover, in order to better understand the mechanism of photoionization-induced broadband DW generated in the Ar-filled SR HC-PCF, we numerically simulated the propagation of ultrashort pulses along the SR HC-PCF using the single-mode unidirectional pulse propagation equation (UPPE) [22-24], the corresponding results are shown in Fig. 3(b). In the simulations, we used the pulses measured by SHG-FROG as the input. The BR and ZS models were used to simulate the fiber loss and dispersion, respectively. The gas ionization was also included in the UPPE, calculated by the Perelomov-Popov-Terent'ev model [25]. The overall agreement between numerical results and experimental results is qualitatively good. According to the magnitude of incident pulse energy, we investigate the influence of the soliton compression (SC) and plasma-driven BS on resonance-induced DW from four stages (marked as the white Roman numbers i, ii, iii and iv), which are separated by white dashed lines.

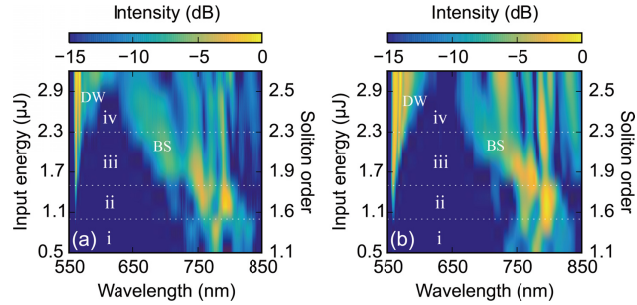

**Figure 3.** Measured (a) and simulated (b) spectral evolutions after propagating a 25-cm-long SR HC-PCF with 24-μm core diameter and 1.4-bar Ar as a function of input pulse energy. In both (a) and (b), DW = dispersive wave, BS = blueshifting soliton.

At low pulse energy levels (from 0.5 μJ to 1 μJ), the pulses undergo the SC process due to the combined effects of SPM and waveguide-induced anomalous dispersion. As shown in Figs. 4(a) and 4(b), we plot the simulated temporal and spectral evolutions in a 25-cm-long Ar-filled SR HC-PCF at input energy of 0.5 μJ. As the propagation distance increases, the peak power of the self-compressed pulses increases rapidly, resulting in a rapid accumulation of plasma [marked as the white circle lines in Fig. 4(b)] due to gas ionization. In Fig. 4(c), we can see that the simulated spectrum (red solid line) at the fiber output is in good agreement with the experimental spectrum (green shaded region). At certain pulse energy levels (from 1 μJ to 1.5 μJ), the SC process leads to a wider pulse spectrum, which results in the frequency to overlap with the linear wave, and then emits the phase-matched DW. At input pulse energy of 1.1 μJ, the self-compressed pulses reach the maximum soliton compression point (MSCP) at the positon of ~7.5 cm. Meanwhile, the pulses obtain the maximum spectral broadening and trigger the phase-matched DW generation, as shown in Figs. 4(d) and 4(e). The simulated output spectrum still agrees well with the measured spectrum [see Fig. 4(f)]. At high pulse energy levels (from 1.5 μJ to 2.3 μJ), the plasma-driven BS not only enhances the energy transfer from the pump light to the DW, but also causes the spectral redshift of the DW spectrum, as shown in Figs. 4(g)-4(i) at input energy of 2 μJ. The central wavelength of the BS is close to the resonance band of the SR HC-PCF, which leads to the phase-matched DW in the long-wavelength region. This has been demonstrated in the recent experiments, and a good explanation is given using the phase-matched condition between soliton and linear wave [17]. In Fig. 4(i), we can observe that the shift of the blueshifting spectrum to the shorter-wavelength region is smaller than the measured spectrum. This is mainly caused by the spectral recoil effect [26]. As the propagation distance increases, the soliton emitting DW gradually loses energy, transferring it to the DW. To conserve the overall energy of the photons, the central frequency of the soliton is shifted to the opposite direction of the DW radiation. Moreover, in the experiment, since the coupling efficiency drops rapidly at high pulse energy levels, the actual pulse energy launched into the Ar-filled SR HC-PCF is less than the value in the simulation, resulting in the spectral recoil effect in the simulation being greater than in the experiment. At a higher pulse energy region (from 2.3 µJ to 3.2 µJ), the plasma-driven BS can further trigger the broadening of DW spectrum to the longer-wavelength side, which leads to an overlap between the blueshifting spectrum and the DW [see Figs. 4(j)-4(l) with an input energy of 2.6 µJ].

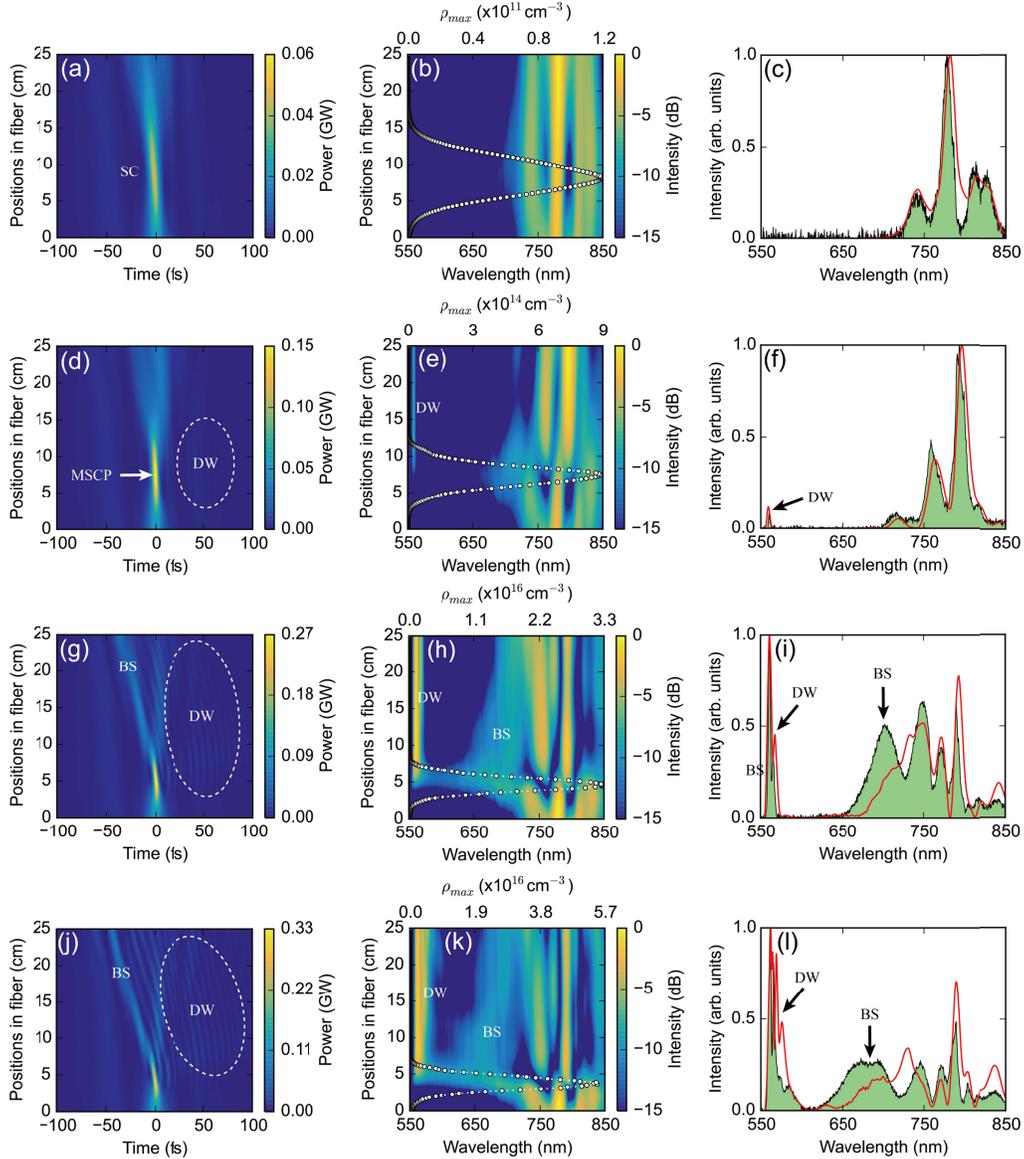

**Figure 4.** Simulated temporal [(a), (d), (g) and (j)] and spectral [(b), (e), (h) and (k)] evolutions in a 25-cm-long Ar-filled SR-PCF with 24-µm core diameter and 1.4-bar gas pressure at different input energies. (a)-(c), (d)-(f), (g)-(i) and (j)-(l) correspond to input pulse energies of 0.5 µJ, 1.1 µJ, 2.0 µJ and 2.6 µJ, respectively. The measured (green shaded region) and simulated (red solid line) normalized spectral intensities for different input energies at the output of the SR HC-PCF are shown in (c), (f), (i) and (l). The white circle lines in panels (b), (e), (h) and (k) point out the maximum plasma density. In (a), SC = soliton compression, in (d), MSCP = maximum soliton compression point.

The phase-matched DW is usually generated in the positive direction of the time axis. In Fig. 5(a), we used a super-Gauss window to filter out the leftmost DW spectral band (marked as red dashed line, called DW-SB1), and the green solid line corresponds to the simulated spectrum at the output of the SR HC-PCF with the input pulse energy of 2.6 μJ. Through the Fourier transform of DW-SB1, the corresponding temporal profile in Fig. 5(b) is located on the right side of the time axis and shows a long pulse duration of ~198 fs at FWHM due to a narrow spectral width. However, the plasma-driven BS located in the anomalous dispersion region has a higher group velocity than input pulses, so that the blueshifting pulses accelerate as the propagation distance increases. This causes the DW radiation to approach the zero point of the time axis, and even appear in the negative direction, as shown in Figs. 5(c)-5(f). The pulse durations corresponding to the DW-SB2 and DW-SB3 are ~171 fs and ~36 fs, respectively. In Fig. 5(h), we also plot the temporal intensity of BS [see Fig. 5(g)], and the pulse duration is about 13 fs. Figure 5(j) shows the complete temporal profile of the output pulses. The main peak is the BS, and its peak position is about 44 fs, consistent with Fig. 5(h). In addition, the left side of the main peak is the existing pedestal of the incident pulses [see Fig. 2(e)], while the oscillations on the right are the DW radiation and residual pump light. The self-compressed pulse duration can be as short as ~6 fs. These results can also be observed in Fig. 4(j).

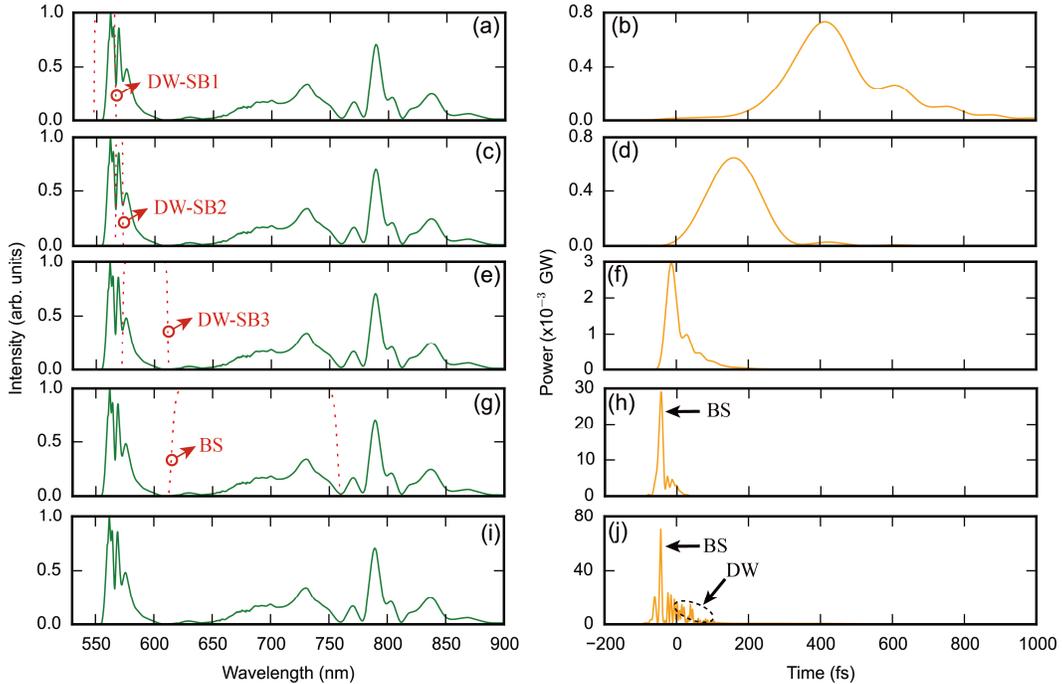

**Figure 5.** The green solid lines represent the simulated spectrum output of a 25-cm-long SR HC-PCF at an input energy of 2.6 μJ. In (b), (d), (f) and (h), the orange solid lines indicate the temporal profiles of the pulses through Fourier transforming the spectral bands (marked as red dashed lines) filtered by super-Gauss windows, and in (j), the temporal profile of the pulse obtained by Fourier transforming the entire output spectrum. In (a), (c) and (e), DW-SB = dispersive wave spectral band.

Figure 6(a) shows the broadband DW spectrum filtered from the output spectrum of Fig. 5(i) through using a super-Gauss window from 550 nm to 610 nm, and the corresponding temporal profile after the Fourier transform is plotted in Fig. 6(b). Although the pulse duration at FWHM is ~30 fs, the pulses exhibit long-decay pedestals on the trailing edge due to the DW radiation generated at different positions in the fiber. We found that after -4634 fs$^2$ group delay dispersion (GDD) and -44827 fs$^3$ third-order dispersion (TOD) compensation, these pedestals can be compressed into the main peak, showing a clean pulse leading edge. The compressed pulse duration is ~29 fs, which is very close to the Fourier transform limit (FTL) of ~24 fs.

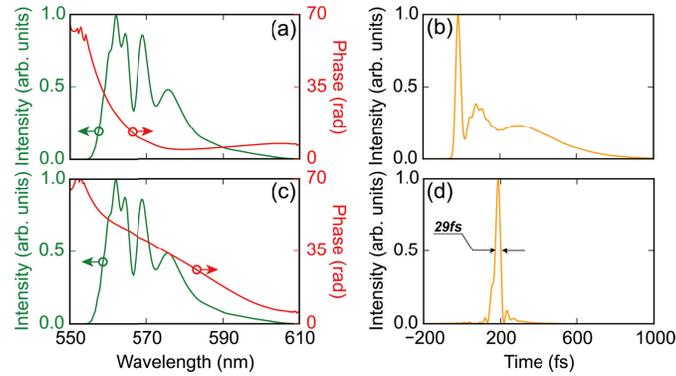

**Figure 6.** (a) The spectrum (green solid line) filtered from the output spectrum of Fig. 5(i) by using a super-Gauss window from 550 nm to 610 nm, and the corresponding phase is marked as red solid line. After GDD and TOD compensation, the spectrum and phase are shown in panel (c). (b) and (d) The temporal profiles of the pulses by Fourier transforming the corresponding spectra.

## 4. Conclusions

In conclusion, we experimentally and numerically demonstrated the photoionization-induced broadband phase-matched DW with several spectral peaks generated in the resonance band of a 25-cm-long Ar-filled SR HC-PCF with core diameter of 24 μm and gas pressure of 1.4 bar. At certain pulse energy levels, the soliton self-compression process can result in a narrow-band DW spectral peak generated near the wavelength of ~550 nm. At high pulse energy levels, the plasma-driven BS gradually moves to the shorter-wavelength region as the propagation distance increases, leading to the broadening of DW spectrum to the longer-wavelength side. Theses experimental observations are well confirmed by the numerical results simulated through well-known UPPE model. Moreover, through using the super-Gauss windows to filter out the DW spectral peaks, we investigated the time-domain characteristics of the DW and better understood the broadband DW generation process. Furthermore, we found that after suitable GDD and TOD compensation, the DW can be compressed together in the time domain, enabling a good pulse quality. The compressed pulse duration is as short as ~29 fs, approaching to the FTL duration of ~24 fs. These experimental results not only offer some useful insights into the resonance-induced DW radiation in gas-filled HC-PCFs, but also provide a simple and effective method for generating the ultrashort light sources in the visible region, which may have many applications in pump-probe spectroscopy.

**Funding:** Zhangjiang Laboratory Construction and Operation Project, grant number 20DZ2210300; National Key R&D Program of China, grant number 2017YFE0123700; National Natural Science Foundation of China, grant number 61925507; Program of Shanghai Academic/Technology Research Leader, grant number 18XD1404200; Strategic Priority Research Program of the Chinese Academy of Sciences, grant number XDB1603; Major Project Science and Technology Commission of Shanghai Municipality, grant number 2017SHZDZX02.